\documentclass[onecolumn,showpacs,preprintnumbers,amsmath,amssymb]{revtex4}
 \usepackage{graphicx}
\usepackage{latexsym}
\usepackage{dcolumn}
\usepackage{stmaryrd}

\usepackage{graphicx}
\usepackage{dcolumn}
\usepackage{bm}

\usepackage{bm}

\begin{document}
\newcommand{\eq}{\begin{equation}}                                                                         
\newcommand{\eqe}{\end{equation}}             

\title{Heat conduction: a telegraph-type model with \\ self-similar 
behavior of solutions} 

\author{ I. F. Barna$^a$ and R. Kersner$^b$}
\address{$^a$ KFKI Atomic Energy Research Institute of the Hungarian Academy 
of Sciences, \\ (KFKI-AEKI), H-1525 Budapest, P.O. Box 49, Hungary, \\ 
 $^b$University of P\'ecs, PMMK, Department of Mathematics and Informatics, 
Boszork\'any u. 2, P\'ecs, Hungary}

\date{\today}

\begin{abstract} 
 For heat flux $q$ and temperature $T$ we introduce a modified Fourier--Cattaneo 
law $q_t+ l \frac{q}{t}= - kT_x .$ The consequence of it
  is a non-autonomous telegraph-type equation.
This model already has a typical self-similar solution which may be written
as product of two travelling waves modulo a time-dependent factor and might 
play a role of intermediate asymptotics.

\end{abstract}

\pacs{44.90.+c, 02.30.Jr}
\maketitle
It is well known, that the heat equation propagates 
perturbations with infinite velocity.
For this contradiction a possible answer is the telegraph equation 
which is "obviously hyperbolic".
In general, one usually forgets about the other fundamental properties of
parabolic heat equations: the existence of self-similar solutions (e.g., the 
Gaussian kernel or fundamental solution) and the attracting 
nature of these special solutions (intermediate asymptotics).

Is is easy to show, that the telegraph equation --see (5)--has 
no self-similar solutions e.g., solutions
of the form $t^{-{\alpha}} f(x/t^{\beta})$ and that even asymptotic 
self-similarity property is lacking:
no solutions of form
$g(t)\cdot f(x/w(t))$ with $g \sim t^{-\alpha}$ and $h \sim t^{\beta}$  for $ t >> 1$.

Since the telegraph equation (possibly with reaction terms) supposed to be 
relevant not only in heat conduction  but  also
in various diffusion processes, the lack in self-similarity might be a 
bad sign for the adequacy of the model.
In addition, in diffusion and heat theory various physical quantities, like the 
fluxes, have to be continuous; therefore the solutions of this equation cannot be "too bad".

According to Gurtin and Pipkin \cite{gurt,jos1,jos2}, the most general 
form of the flux in  linear heat
conduction and diffusion related to the flux $q$  expressed in one space dimension
via an integral over the history of the temperature gradient
\begin{equation}\label{1}
  q = -\int_{-\infty}^{t}Q(t-t')\frac{\partial T(x,t')}{\partial x}dt'
\end{equation}
 where $Q(t-t')$ is a positive, decreasing relaxation function that tends to zero
as $t-t' \rightarrow  \infty $ and $T(x,t)$ is the temperature distribution.

There are two notable relaxation kernel functions: if $Q(s) = k\delta(s)$ where $\delta(s)$ is
a Dirac delta "function", then
\begin{equation}\label{2} 
q= -k\frac{\partial T(x,t)}{\partial x}
\end{equation}
is the Fourier law.

If we define the kernel as
$Q(s) =  \frac{k}{\tau}e^{-s/\tau} $ where
$s = t-t'$ and $k$ is the constants of the effective thermal conductivity 
we get back to the well-known
Cattaneo \cite{cat} heat conduction law which
reads:
\begin{equation}\label{3}
 \tau \frac{\partial {q}}{\partial t} + q = - k\frac{\partial T(x,t)}{\partial x} .
\end{equation}

There is a third kind of relaxation kernel, the "Jeffrey-type", which applies both
$Q(s) =  k_1\delta(s) +  \frac{k_2}{\tau}e^{-s/\tau}. $

 The energy conservation law
\begin{equation}\label{4}
 \frac{\partial q}{\partial x} = -\gamma\frac{\partial T(x,t)}{\partial t},
\end{equation}
where  $\gamma$ is the heat capacity, gives the heat equation with (1) and the
 telegraph equation with (2):
\begin{equation}\label{5}
\frac{\partial^2 T(x,t)}{\partial t^2} + \frac{1}{\tau} \frac{\partial T(x,t)}{ \partial t}
= c^2\frac{\partial^2 T(x,t)}{\partial x^2}, 
 \end{equation}
where $c = \sqrt{k/\tau\gamma} $ is the propagation velocity of the transmitted heat wave. 
The flux $q$ satisfies the same equation.

The thermal diffusivity $\kappa = k/\gamma$ can be defined as the ratio of the
effective thermal conductivity $k$ and the heat capacity $\gamma$. This 
equation describes a physical process
with a well defined relaxation time $\tau$.

The telegraph equation  (\ref{5}) can be derived in various transport systems.
First it was derived by Kirchhoff to describe the voltage on a realistic electrical transmission
line with distance and time, this derivation can be found in any  textbook 
on electrodynamics. Later Goldstein \cite{gold} derived the telegraph equation from a
generalized random walk for diffusion.
Okubo \cite{okubo}  summarized the knowledge of the telegraph equation till 1971 and
presented a new  ad-hoc derivation for turbulent mixing his starting point 
being the Navier-Stokes system. In all these derivations the $\tau$ 
relaxation time had a well-established physical meaning 
presenting the time-scale of the physical process. Joseph and Preziosi collected basically
all the relevant works done on heat waves connected with the 
telegraph equation till 1990 \cite{jos1,jos2}.  
On the other side the telegraph equation was intensively investigated 
from the mathematical point of view as well \cite{othm}. Different kind of properties of these 
models can be analyzed as in \cite{kers}. \\ 
In the present communication we introduce a new kernel which somehow interpolates 
the Dirac delta  and the exponential kernel having the main properties 
of both. The $ Q(s) = 1/s^l$ is such a function: it is singular at the origin and
has a short range of decay for $ l>1$.
Let's consider the following relaxation kernel
\begin{equation}\label{6}
Q(t-t') = \frac{k\tau^l}{(t-t'+\omega)^l}
 \end{equation}
where $k$ is the effective thermal conductivity, $\tau$ is a relaxation time and
$l>1$ is a parameter,  $-t' + \omega$ is just a time shift which is needed
to regularize the expression.

Using the general form of heat flux (\ref{1}) we get
\begin{equation}\label{7}
q = -\int_{-\infty}^{t} \frac{k\tau^l}{(t-t'+\omega)^l} \frac{\partial T(x,t)}{ \partial x}dt'.
\end{equation}

One has
\begin{eqnarray}\label{8}
\frac{ \partial q}{\partial t} =  -k\left(\frac{\tau}{\omega}\right)^l
\frac{\partial T(x,t)}{\partial x} +  \nonumber \\
l\int_{-\infty}^{t} \frac{1}{t-t'+\omega}  \frac{k\tau^l}{(t-t'+\omega)^l}
 \frac{ \partial T(x,t')}{\partial x}  dt'.
\end{eqnarray}
A {\it formal} application of integral mean theorem to the second term on the right
hand side and the definition of $q$ leads to a new phenomenological law

\begin{equation}\label{9}
\frac{ \partial q}{\partial t} =  -k\left(\frac{\tau}{\omega}\right)^l
 \frac{ \partial T(x,t)}{\partial x}
-  \frac{l}{t-t''+\omega} q.
\end{equation}
The additional energy conservation law is still (\ref{4}) and from this and the last equation
we obtain
\begin{eqnarray}\label{10}
 \frac{\gamma}{k}   \left( \frac{\omega}{\tau} \right)^l \frac{\partial^2
T(x,t)}{\partial t^2} + \nonumber \\
  \frac{\gamma}{k}   \left( \frac{\omega}{\tau} \right)^l \frac{l}
{t-t''+\omega}  \frac{\partial T(x,t)}{\partial t} =   
\frac{\partial^2 T(x,t)}{\partial x^2}.
\end{eqnarray}

For a better transparency let's call $\epsilon = \frac{\gamma}{k}
\left(\frac{\omega}{\tau}\right)^l $
and $a = \frac{\gamma}{k} \left(\frac{\omega}{\tau}\right)^l \cdot l$
The physical meaning of $\epsilon$ is still the thermal diffusivity multiplied
by a scaling constant which is the renormalized relaxation 
time (the ratio of an ordinary time shift
$\omega$ and a well defined relaxation time $\tau$).  The exponential $l$ is a real number which
describes the non-locality in time which we may call memory
effects of the heat conduction phenomena. Larger $l$ means
shorter memory.
The physical meaning of $a$ is approximately the thermal diffusivity multiplied
by another time-scaling factor. In the following we will see that the role of $a, \epsilon$
or $l$ will be crucial in the structure of the solutions.
At last we introduce a new time variable $ t = t-t''+ \omega.$
Now our telegraph-type equation reads
\begin{equation}\label{11}
\epsilon\frac{\partial^2 T(x,t)}{\partial t^2} +
\frac{a}{t} \frac{\partial T(x,t)}{\partial t} = \frac{\partial^2
T(x,t)}{\partial x^2}.  
\end{equation}
Note, that the $a/t$ factor appearing in front of the first time derivate 
makes the equation time-reversible, which cannot be true for diffusion or heat 
propagation processes, at the same time the $a/t$ factor makes the 
equation irregular at the origin. To avoid these problems, we may shift the pole to a negative time value - in practical 
calculations using the $a/(t+\tau)$ where $\tau$ still can be any kind of relaxation time with well-founded physical interpretation. Physically it is clear, if a process has a 
well-defined time-scale than the reverse process cannot run back in time more than the physically 
relevant time.      
Now, in this sense, for positive time $(t>0)$ we may use the equation 
to describe diffusion-like processes. 
Our deeper investigation clearly shows, that no other exponent of $t$ as 
one can have self-similar solution. 

If we consider (\ref{11}) as a non-linear wave equation we may investigate 
wave properties like dispersion phenomena. 
Inserting the standard plain wave approximation $T(x,t) = e^{i(\tilde{k}x + \tilde{\omega}t)}$ into
(\ref{11}) the dispersion relation and the attenuation distance can be obtained.
These are the followings:
\begin{equation}
v_p =  \frac{\tilde{\omega}}{Re(\tilde{k})} = \sqrt{\frac{2}{\epsilon}} {\tilde{\omega}}
\left( 1+ \sqrt{1 + \left( \frac{l}{t} \right)^2 } \right )^{-1/2} \hspace{0.3cm}
\tilde{\alpha} = \frac{1}{Im(\tilde{k})} = \frac{2t}{\epsilon l}\frac{1}{v_p}  . 
\end{equation} 
Our telegraph-type equation is time dependent, hence both the dispersion relation and the
attenuation distance have time dependence.
Note, that $v_p$ has a  very weak time-dependence, basically only till $t \le l$.
As an other interesting point is that the  phase velocity does not depend on the angular velocity which
is the same as for the ideal wave equation, so our equation has no dispersion.
So in this sense our equation is very similar to the wave-equation, which is hyperbolic.
The properties of the attenuation distance is even more interesting,
it is divergent in time and has a $1/\tilde{\omega}$ angular frequency.
However if we let the angular frequency and the time to go infinite with the same speed than the
attenuation distance has a strong decay. Which is like the skin-effect when high frequency electrons
can only propagate on the surface of a metal.


We are looking for solution of (\ref{11}) of the form
\begin{equation}
T(x,t)=t^{-\alpha}f\left(\frac{x}{t^\beta}\right):=t^{-\alpha}f(\eta).
\end{equation} 
The similarity exponents $\alpha$ and $\beta$ are of primary physical importance since 
$\alpha$  represents the rate of decay of the magnitude $T(x,t)$, 
while $\beta$  is the rate of spread 
(or contraction if  $\beta<0$ ) of the space distribution as time goes on.  
Substituting this into (\ref{11}) we have
\begin{eqnarray}\label{12}
f''(\eta) t^{-\alpha -2} [\epsilon\beta^2 \eta^2] + \nonumber \\
f'(\eta) \eta t^{-\alpha -2} [\epsilon\alpha\beta  -
\epsilon\beta(-\alpha-\beta-1)  - \beta a] + \nonumber \\
f(\eta) t^{-\alpha -2} [-\epsilon\alpha(-\alpha-1) - a\alpha] =
f''(\eta) t^{-\alpha -2\beta}
\label{3},
\end{eqnarray}
where prime denotes differentiation with respect to $\eta.$

One can see that this is an ordinary differential equation(ODE)
if and only if $\alpha + 2 = \alpha + 2\beta$ ({\it {the universality relation}}).
So it has to be 
\begin{equation}
\beta =1  
\end{equation} 
while $\alpha$ can be any number. The corresponding ODE
we shall deal with is
\begin{equation}\label{13}
    f''(\eta) [\epsilon\eta^2-1 ] +
f'(\eta) \eta(2\epsilon\alpha  + 2\epsilon -a) +
f(\eta) \alpha (\epsilon \alpha+\epsilon- a) = 0.
\end{equation}

\begin{figure}*
\scalebox{0.65}{
\rotatebox{0}{\includegraphics{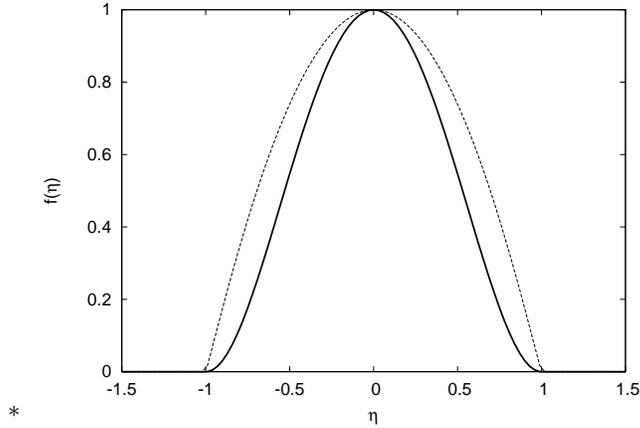}}}
\vspace*{0.4cm}
\caption{Eq. (21) thick solid line is for $l = 6.2$ and the thin dashed line is for $l = 4.1$. }	 
\label{elso}       
\end{figure}

In pure heat conduction-diffusion processes--no sources  or sinks--the heat mass is
conserved: the integral of $T(x,t)$ with respect to $x$ does not depend on time $t$.
For $T(x,t)$ this means
\begin{equation}
\int T(x,t) dx=t^{-\alpha} \int f(\frac{x}{t})dx=t^{-\alpha +1}\int f(\eta)d\eta=const
\end{equation}
if and only if $\alpha=1.$
We are going to investigate this case only. Plainly (\ref{14}) can be written as
\begin{equation}
(\varepsilon f\eta^2-f)''=a(\eta f)'
\end{equation}
which after integration and supposing $f(\eta_0)=0$ for some $\eta_0$ gives
\begin{equation}\label{14}
    \frac{df}{f}=\frac{a\eta d\eta}{\varepsilon \eta^2-1}.
\end{equation}
From this equation we can obtain two qualitatively different solutions. 
The one which is globally bounded and positive
in the domain $\{(x,t): 1-\varepsilon \eta^2 >0 \}$ and has the form \begin{equation}
f=(1-\varepsilon \eta^2)^{\frac{a}{2\varepsilon}-1}_+
\end{equation}
 where $(f)_+=\max (f,0).$
See Fig. (\ref{elso}). 

The corresponding self-similar solution is

\begin{equation}\label{15}
    T(x,t)=\frac{1}{t}\left(1-\varepsilon \frac{x^2}{t^2}\right)^{\frac{a}{2\varepsilon}-1}_+
\end{equation}

\begin{figure}*  
\scalebox{0.35}{
\rotatebox{-90}{\includegraphics{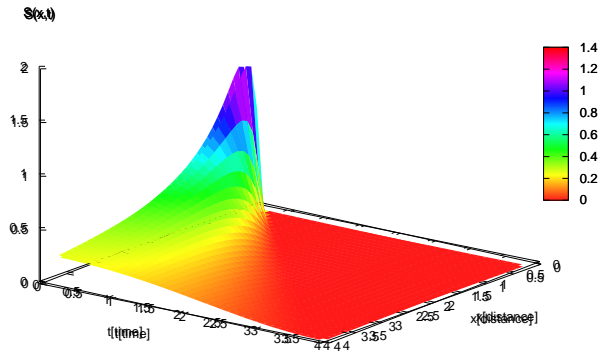}}}
\caption{The solution (\ref{15}) for the parameter $l=6.2$.  }	
\label{kettes}       
\end{figure}

This solution is positive in the cone $t^2> \varepsilon x^2$ and is zero 
outside of it, see Fig. (\ref{kettes}). Note, that only the $x>0$ and $t>0$ 
quarter of the plane is presented, because 
of it has physical relevance. 

On the $(x,t)$ plane there are two fronts $x(t)=\pm \frac{t}{\sqrt{\varepsilon}}$ 
separating these domains. Because the function $T(x,t)$ not always has 
continuous derivatives entering to (\ref{11}) we have to make clear what we 
mean under "solution". Having in mind the physical background, we ask the 
continuity of $T_t, T_x, q_t$ and $q_x$ so
that in (2) and (3) all functions were continuous.
In our case this means that
\begin{equation}\label{16}
    \frac{a}{2\varepsilon}-1=\frac{l-2}{2}
\end{equation}
 has to be greater than $1$, i.e. $a/\varepsilon=l>4,$  which we shall suppose further on.
 If the second derivatives are not continuous ($4<l\leq 6$) we 
understand the solution in the sense of distributions. If $l>6$ the solution is classical.
On Fig. 1. we compare the solutions with $l=4.1$ and $l=6.2$. the thick 
solid line represents the solution for $ l= 6.2$ and the thin dashed 
line however shows the solution for $l=4.1$. 

  {\bf Remark 1.} The solution (\ref{15}) is of {\it source-type} 
i.e. $\lim_{t\rightarrow 0}T(x,t)= K\delta(x),$ where $\delta$ is the 
Dirac measure, $K>0$. One can calculate the second initial 
condition $\lim_{t\rightarrow 0}T_t(x,t)$ too.

  {\bf Remark 2.} One can write (\ref{15}) in the form of {\it product} of two 
traveling waves propagating in opposite direction (divided by a time-factor):
\begin{equation}
T(x,t)=\frac{1}{t^{l-1}}(t-\sqrt{\varepsilon}x)_+^{\frac{l}{2}-1}(t+\sqrt{\varepsilon}x)_+^{\frac{l}{2}-1}
\end{equation}
  which is a new-type of purely hyperbolic wave; the typical 
solution of the wave equation is the {\it sum} of two such waves: $g(x-ct)+g(x+ct)$.

It is known that an another possible answer to contradiction 
connected with the infinite speed of propagation is the nonlinear Fourier 
law ($\tau=0, k=k_0 T^{m-1}$ in (2)) which leads 
to a nonlinear heat equation

  \begin{equation}\label{17}
    T_t=(T^m)_{xx}, \quad m>1.
  \end{equation}
  In \cite{zk} Zeldovich and Kompaneets have found the fundamental 
solution $T_1$ of this equation which we write in the following  form:
  \begin{eqnarray}\label{18}
    T_1^{m-1}=t^{-\alpha (m-1)}\left(A^2-B^2x^2t^{-2\beta}\right)_+=  \nonumber  \\
\frac{1}{t}(At^\beta-Bx)_+(At^\beta+Bx)_+,
  \end{eqnarray}
  where $A$ is constant  and 
\begin{equation}
\alpha=\beta=\frac{1}{m+1},\quad B^2=\frac{m-1}{2m(m+1)}.
\end{equation}
 One can see that this solution has bounded support in $x$ for any $t>0$ 
which is a hyperbolic property. Using comparison principle for such 
equations one can show this finite speed property for 
any initial condition having compact support. However, the fronts are 
not straight lines: $x(t)=\pm \frac{A}{B}t^\beta$, 
$\beta <1$ so the speed
 of propagation $\dot{x}(t)$ goes to zero if $t$ goes to infinity. One can 
also see that $T_1$ is of source-type: $T_1(x,0)=K_1\delta(x).$
 
 The most intrinsic property of $T_1$ is that it plays the role of 
{\it intermediate asymptotic}: any solution of
 (\ref{17}) corresponding to the initial datum $t(x,0)$ with $\int 
t(x,0)dx=K_1$ converges to $T_1$ as $t\rightarrow \infty.$ This was 
conjectured earlier but was shown only in 1973 by Sh. Kamin, see \cite{ka}.
 
 It would be important and interesting to understand whether or not 
our special solution $T(x,t)$ had this attractor property. If "yes", 
in what sense: we recall that there is a second initial condition too. \\ 

{\it{In summary. -}}
We introduced a new phenomenological law for heat flux which in some sense
"interpolates" between Fourier and  Cattaneo laws. The consequence of it is a
 non-autonomous model, a telegraph-type partial 
differential equation. It already has, unlike the classical telegraph equation, 
self-similar solutions, the presence of which 
is desirable in the theory of heat propagation free from sources and absorbers.

One of us (R.K.) would like to thank Prof. P. Rosenau for his stimulating 
discussion. 

                                                                  

\begin{thebibliography}{a}  

\bibitem{gurt} M.E. Gurtin and A.C. Pipkin,
Arch. Ration. Mech. Anal. {\bf{31}}, 113 (1968).

\bibitem{jos1} D.D. Joseph and  L. Preziosi,
Rev. Mod. Phys. {\bf{61}}, 41 (1989).

\bibitem{jos2} D.D. Joseph and  L. Preziosi,
Rev. Mod. Phys. {\bf{62}}, 375 (1990).

\bibitem{cat} C. Cattaneo,  Sulla conduzione del calore 
{\it{Atti. sem Mat. Fis. Univ. Modena }}
{\bf{3}}, 83 (1948).

\bibitem{gold} S. Goldstein,
Quart. J. Mech. and Appl. Math.
{\bf{4}}, 129 (1951).

\bibitem{okubo}A. Okubo, {\it{Application of the Telegraph Equation to Oceanic 
Diffusion: Another Mathematical Model}} Chesapeake Bay 
Institute, The John Hopkins University, Technial 
Report 69, N00014-67-A-0163-0006 NR 083-016 \\ 
http://dspace.udel.edu:8080/dspace/handle/19716/1439.
 
\bibitem{othm} H.G. Othmer, S.R. Dunbar and W. Alt, J. Math. Biol. {\bf{26}}, 
263 (1988).

\bibitem{kers} B.H. Gilding and  R. Kersner, {\it{Travelling Waves in 
Nonlinear Diffusion-Convection Reactions,}} Progress 
in Nonlinear Differential Equations and Their Applications, 
Birkh\"auser Verlag, Basel-Boston-Berlin, 2004, ISBN 3-7643-7071-8.

\bibitem{ka} Sh. Kamin, Israeli Journal of Maths, {\bf{14}}, 76 (1973).
 
\bibitem{zk} Y.B. Zeldovich and A.S. Kompaneets, {\it{Collection Dedicated to 
the 70th Birthday of A.F. Joffe}}, Izdat. Akad. Nauk SSSR 1950, p.61.

\end{thebibliography}
\end{document}